# The network structure of visited locations according to geotagged social media photos

Christian Junker[1], Zaenal Akbar[2], Martí Cuquet[2]

[1] Fanlens.io, Baumkirchen, Austria
christian@fanlens.io
[2] Universität Innsbruck, Technikerstraße 21a, 6020 Innsbruck, Austria
{zaenal.akbar, marti.cuquet}@sti2.at

**Abstract.** Businesses, tourism attractions, public transportation hubs and other points of interest are not isolated but part of a collaborative system. Making such collaborative network surface is not always an easy task. The existence of data-rich environments can assist in the reconstruction of collaborative networks. They shed light into how their members operate and reveal a potential for value creation via collaborative approaches. Social media data are an example of a means to accomplish this task. In this paper, we reconstruct a network of tourist locations using fine-grained data from Flickr, an online community for photo sharing. We have used a publicly available set of Flickr data provided by Yahoo! Labs. To analyse the complex structure of tourism systems, we have reconstructed a network of visited locations in Europe, resulting in around 180,000 vertices and over 32 million edges. An analysis of the resulting network properties reveals its complex structure.

## 1 Introduction

The current ubiquity of digital and hyperconnected activities generates an ever-growing amount of available data. Coupled with the increasing ability to process, link, analyse and exploit them, it is producing a radical impact in our society and how individuals and organisations function and interact.

This new reality of data-rich environments is posing novel challenges and opportunities emerge that are not only technical [1], but also expand into the economic, social, ethical, legal and political fields. Some examples are an increased efficiency and innovation speed, the appearance of new business models, raising concerns on data quality, reliability and trust as well as privacy, protection and accountability issues, among others [2]–[4].

As a result, businesses and economic sectors are adapting to this new reality. Research is also quickly embracing the potential of using and analysing this expanding number of data sources. The study of complex and collaborative systems can also substantially benefit from these large amounts of evolving data. Indeed, the use of big data, which share the large scale (volume), complexity (variety) and dynamics (velocity) properties of complex systems [5], enhanced by the innovative potential of open data [6], and machine learning, data mining and natural language processing tools, among others, set an ideal framework for a data-driven approach to the study of collaborative networks of autonomous entities cooperating to achieve a common or compatible goal.

In some applications, this approach is proving very productive, e.g. in air traffic management [7], face-to-face behavioural networks in human gatherings [8], and movements of farmed animal populations [9]. Data-centric fields, of which these are examples, provide an empirical framework where advances in network science, and particularly in collaborative networks, can be tested. The complex structure of actors and their relations is particularly relevant in socioeconomic systems. This has triggered a long history of interdisciplinary collaboration between network science and fields such as computational sociology [10], transportation systems [7], [11], economy [12], and also that of collaborative networks, which is increasingly benefiting from data-driven approaches [13], [14].

A field with a particularly big potential to benefit from intensive data-driven network research but that has still been hardly explored is the tourism sector. Some early studies include a characterisation of the worldwide network of tourist arrivals at the country level [15] and of touristic destinations [16]–[19]. The tremendous increase in the abundance of data sources in the tourism sector is boosting a new data intensive approach [20]. Examples of sources are online bookings, the process of tourists informing themselves before the travel, and the sharing of their experiences during and after it via social media. Some examples are the use of geotagged data of tourists to show the destination preference and the hotspots in a city [21], analyse sentiment by neighbourhoods [22], describe city and global mobility patterns [23], [24] and predict taxi trip duration [25]. Social media data may be used as a source to reveal business and points of interest relationships and thus open the ground for collaborative value-creation.

In this paper, we reconstruct a European network of locations visited by tourists using fine-grained data from Flickr, an online community for photo sharing. We have used a publicly available set of Flickr data provided by Yahoo! Labs [26]. The network design relies on the use of collaboratively contributed data by users: The locations where photos were taken make the nodes of the network, and are connected if at least two different Flickr users took a photo in both locations.

Social media networks in particular contain salient data that highlights real-world behaviour patterns of their users. Due to these properties, these networks can act as the catalyst for the reconstruction of complex, possibly multilayered connections in seemingly unrelated networks. This study shows the feasibility and potential of using social media data in the collaborative networks field, and reconstructs the relationships between relevant places for tourists with the aim to contribute to a better understanding of what constitutes the central and most relevant points of interest. Further, results of the study could make a significant contribution in assisting the design of collaborative networks of city entities in the face of tourism, be they businesses, landmarks, attractions, public transport authorities or others. Finally, it lays the ground for future research to reconstruct multiplex location networks, where each of the layers corresponds to different segmentation of users, such as locals and tourists or by country of origin.

This paper is organised as follows. In Section 2 we present the network of locations in Europe visited by Flickr users. First, the YFCC100M dataset is briefly presented and discussed. We then outline the methodology used to prepare the network from this dataset, and finally proceed to the network analysis. Section 3 discusses the results and we conclude with some remarks in Section 4.

## 2 Network reconstruction and analysis

### 2.1 Flickr dataset

The Yahoo Flickr Creative Commons 100 Milion Dataset (YFCC100M) [26], released in 2014, is a public dataset of 100 million media objects uploaded to Flickr, a social image and video hosting website. Almost all of its contents cover the period between 2000 and 2014. The dataset is very rich in metadata, enabling a large variety of applications. Since its release, it has been used in a variety of contexts, such as photo clustering [27], multimodal learning [28], situation recognition [29], trajectory recommendation [30], and tag recommendation.

The metadata contained in the dataset, aside from Flickr-related data such as a photo identifier and the user that created it, include tags used by users to annotate it (68 million objects have been annotated), camera used, time when the photo was taken and when it was uploaded, location and license. For this paper, only the metadata related to the geolocalisation has been used, although future work would largely benefit from consideration of at least tags and timing, to enable e.g. a dynamic analysis of the network. In total, 48 million objects are annotated with the geolocalisation of the object, and the most prominent cities represented in the dataset are London, Paris, Tokyo, New York, San Francisco and Hong Kong [26]. **Error! Reference source not found.** shows the locations of all those photos, linked as described below in Section 2.2.

### 2.2 Collaborative network reconstruction

To analyse the complex structure of tourism systems, we have used Apache Spark for the pre-processing of the YFCC100M dataset and converted it into a GraphX graph to construct a network of locations visited by users of Flickr. In this undirected weighted network, a vertex corresponds to the geolocation of a media object in the YFCC100M dataset as specified by the latitude and longitude fields. We used a precision of $10^{-3}$ degrees both in latitude and longitude, which at 45° of latitude roughly corresponds to 111 meters of latitude and 79 meters of longitude. In practice, this means that media objects show up as the same vertex if they are on the same street or neighbourhood. The network is represented by a graph $G = \{V, E\}$, where $V$ is the set of vertices and $E$ is the set of edges. Two vertices $u$ and $v$ are connected by an undirected edge $(u,v)$ if at least two different users have a media object in the two locations corresponding to such vertices (i.e., both visited the two locations). Reconstructing the network without this constraint leads to tremendous noise, i.e. spurious connections between singular points of interest. The weight $w_{uv}$ of an edge $(u,v)$ is the number of users that visited locations $u$ and $v$. The resulting network for Europe has $N$=178,661 vertices and $M$=32,753,756 edges.

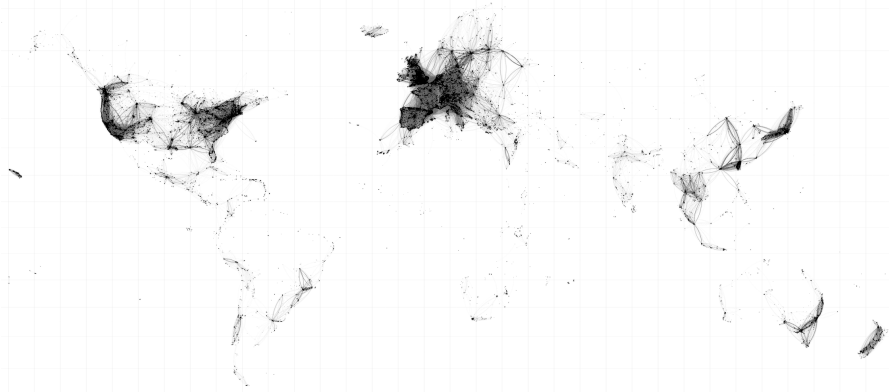

**Fig. 1.** Global overview of the geolocalised photos of the YFCC100M dataset [26]. Locations where photos were taken have been linked following the method described in Section 2.2. For clarity, only links connecting locations separated by less than 10 degrees are displayed.

### 2.3 Network analysis

The network of visited locations in Europe consists of one giant connected component of 174,699 nodes, accounting for 97.8% of the total, and 1,575 other small components of sizes ranging from 2 (most of them) to 29.

The degree $k_u$ of vertex $u$ is the number of edges attached to the vertex. In the present network, it is the number of locations that were visited by the same users that visited a given location, and thus indicates what are the hotspots in the city or region. One of the most important characteristics of real-world networks is their degree distribution $p_k$ [31]: the probability that a randomly chosen vertex has degree $k$. In the binomial random graph model, each of the $\binom{N}{2}$ pairs of vertices holds an edge with a certain probability $p = \langle k \rangle/(N\text{-}1)$, with $\langle k \rangle$ the average degree. For large graphs with $N \rightarrow \infty$, its probability distribution tends to a Poisson distribution, $p_k = \langle k \rangle^k e^{-\langle k \rangle}/k!$. Real-world networks, on the other hand, typically have a larger number of nodes of high degree, and follow a distribution that decays as a power law, $p_k \sim k^{-\theta}$, rather than exponentially [31]. In our case of the European network of locations, it decays as a power law with exponent $\theta$=1.34, as shown in Fig. 2.

In a weighted network, the degree alone is not enough to characterise the relation between a node and the rest of the network. Indeed, each edge (*u*,*v*) in our network is weighted according to the number of users that visited both endpoints, *u* and *v*, of the edge, with each user adding 1 to the weight $w_{uv}$. The range of weights goes from 2 to 944 users with an average of 303.78. The probability distribution of weights, shown in Fig. 2, follows again a power law $p_w \sim w^\gamma$ with exponent $\gamma=2.89$.

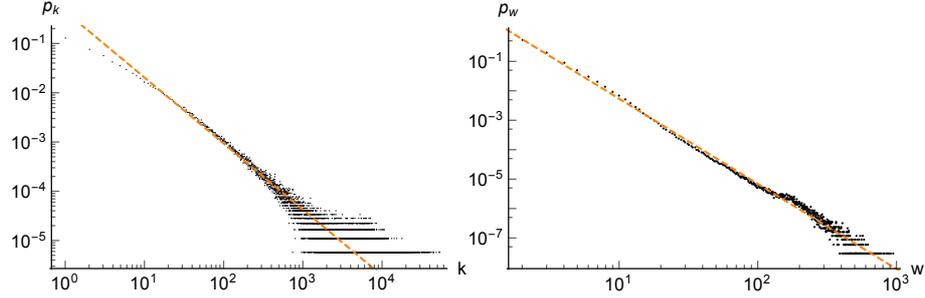

**Fig. 2.** Left: Log-log plot of the degree distribution of the network, with a power law decay $p_k \sim k^{-\theta}$, where $\theta=1.34$. Right: Log-log plot of the weight distribution of the network, with a power law decay $p_w \sim w^{-\gamma}$, where $\gamma=2.89$.

We also analyse if there is a correlation on how locations are linked to each other in terms of the location degree. Typically, social networks tend to show assortative mixing, i.e. nodes tend to be connected to other nodes of similar degree. On the contrary, economic, technological and biological networks tend to show disassortative mixing, where nodes of high degree tend to connect to nodes of low degree [32]. To examine the assortativity of our network, we consider the average degree of the neighbours of a node with degree *k*,

$$\langle k_{nn} \rangle = \sum_{k'} k' \, p'(k'|k) \, , \quad (1)$$

where $p'(k'|k)$ is the conditional probability that an edge leaving a node of degree *k* leads to a node of degree *k'*. This probability is proportional to $k'p_{k'}$ if it is independent of *k* [33]. Fig. 3 shows the $\langle k_{nn} \rangle$ distribution for our network and indicates a rather weak degree-degree correlation. We thus computed the Pearson correlation coefficient of the degrees at the ends of an edge,

$$r = \frac{1}{\sigma_q^2} \sum_{j,k} jk(e_{jk} - q_j q_k) \, , \quad (2)$$

which is in the range $-1 \leq r \leq 1$. Here $q_k$ is the remaining degree distribution,

$$q_k = \frac{(k+1)p_{k+1}}{\sum_j j p_j} \, , \quad (3)$$

$e_{jk}$ is the joint probability distribution of the remaining degree for the two vertices of a same edge, and $\sigma_q^2$ is the variance of $q_k$ [32]. In our network, $r = -2.36 \times 10^{-6}$, showing no assortative mixing and confirming the results in Fig. 3.

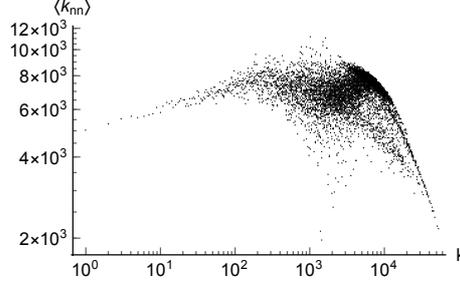

**Fig. 3.** Log-log plot of the average neighbour degree with respect to the node degree. It indicates no assortative mixing, confirmed by a correlation coefficient of $r = -2.36 \times 10^{-6}$.

## 3 Discussion

The work presented in this paper shows how to use the YFCC100M dataset to reconstruct a network of locations visited by Flickr users. The resulting network of around 180,000 vertices and over 32 million edges, comprising all locations in Europe with a granularity at the street/neighbourhood level, displays a complex structure with a scale-free degree and weight distribution, in line with other social, economic and technological networks [34]. An analysis of degree-degree correlations, however, shows no assortative mixing, as opposed to different results in other real-world networks [32], and thus further analysis is recommended that take into account the edge weights and node strengths, as well as exploring the clustering properties.

The increasing data richness of activities associated with tourism activities, especially from the social media domain, exemplified by the present study, make it a highly promising testbed for the study of collaborative networks in the tourism sector. Future steps that would greatly assist in the characterisation and potentiate innovation in collaborative tourism would be the identification of communities and motifs in the network. To this aim, the first requirement is to link the coordinates of the dataset with points of interest, such as local businesses, landmarks or transportation hubs, as has been done in other works. Additionally, smaller networks with a higher detail resolution can be readily obtained with our methodology, enabling the comparison between different cities and possibly revealing different ecosystem patterns.

## 4 Concluding remarks

This study shows the feasibility and potential of using social media data in the collaborative networks field, to link local business, landmarks and other points of interest based on social media users visiting them. It lays the ground for further data-driven studies that make use of the richness of the metadata of similar sources, aside from the geotagging, that allow for future research on multilayered collaborative networks. In that case, different layers could correspond to e.g. countries of origin of the users and assist in the segmentation of users via e.g. community detection, and a better understanding of the collaborative possibilities of tourism.